%% file: main.tex
\documentclass[a4paper,UKenglish,cleveref, autoref, thm-restate]{article}
\usepackage{geometry}
\input{preamble}

\begin{document}
\title{How to Compose Shortest Paths}\author{Jade Master}\date{}
\maketitle
\begin{abstract}\noindent The composition problem for shortest paths asks the following: given shortest paths on weighted graphs M and N which share a common boundary, find the shortest paths on their union. This problem is a crucial step in any algorithm which uses the divide and conquer method to find shortest paths. This extended abstract details how this problem may be understood categorically. Finding shortest paths is represented by a functor and the composition problem asks to find the value of this functor on a pushout using the values of the functor on the components. Furthermore, we present an algorithm which solves the composition problem for shortest paths. When implemented in Python, this algorithm reduces the computation time for finding shortest paths by relying on precompilation. 
\end{abstract}

\section{A Composition Problem}
% The shortest path problem for a weighted graph asks to find the shortest path between a given pair of nodes. 

% Algorithms which solve the shortest path problem typically have quadratic complexity in the number of nodes.
% To redu
Divide and conquer is an effective method for reducing the computation time of many graph algorithms. With this strategy, a problem may be broken up into subdomains, the problem is solved on the subdomains, and then joined together to obtain the final solution. This last step of recombination is the main topic of study for this paper and may be phrased in categorical terms. Other work has studied this recombination step in a similar way for the reachability problem on Petri nets \cite{stephens}. We work in the following abstract setting
\[ \begin{tikzcd}\Set \ar[r,"L",bend left,pos=.4] \ar[r,phantom,"\bot"] & C \ar[r,"F"] \ar[l,"R",bend left,pos=.6] & D
\end{tikzcd}\] where $C$ is a category whose objects are systems broadly construed. The left adjoint $L$ is understood as the "discrete system" functor and the right adjoint is the "underlying set functor". The functor $F \maps C \to D$ represents a computational problem with $D$ representing some category of solutions. This abstract setting is the same which is used in the theory of structured cospan double categories \cite{baezcourser}. The adjoint pair $L \dashv R$ produces a symmetric monoidal double category whose morphisms represent open systems. When $F$ preserves colimits, it lifts to a symmetric monoidal double functor between structured cospan double categories. Functoriality of this double functor means that $F$ may be built from smaller components. However, knowing that $F$ may be built compositionally and actually finding a way to do it are very different matters. In this paper we aim to bridge the gap between theory and practice by providing an algorithm for this composition in the case when $F$ finds the shortest paths on a weighted graph.  In this setting, a pushout\[
\begin{tikzcd}
    M+_{LX} N 
    \arrow[dr, phantom, "\lrcorner", very near start]
    & N \ar[l,"i",swap]  \\
    M \ar[u,"j"]
    & LX \ar[l] \ar[u]
\end{tikzcd}\]
in the category $C$ represents the gluing of a system $M$ with a system $N$ along a set of boundaries $X$. We may now state the crucial step of the divide and conquer method in categorical terms.
\begin{defn}
The \define{composition problem} for $F$ asks: given $F(M)$ and $F(N)$ as inputs, find $F(M+_{LX} N)$.
\end{defn}
\noindent In this paper, $C$, $L$, $R$, and $F$ will be chosen as follows.
\begin{defn}
Let $[0,\infty]$ be the semiring of positive real numbers including infinity. Addition is given by $\min$ and multiplication is given by $+$. A $[0,\infty]$-graph with vertices $X$ is a function $ M \maps X \times X \to [0,\infty]$. For a function $f\maps X \to Y$ and a matrix $M\maps X \times X \to [0,\infty]$, the pushforward matrix $f_*(M) \maps Y \times Y \to [0,\infty]$ is defined by 
\[f_*(M)(i,j) = \sum_{(a,b) \in (f \times f)^{-1}(i,j)} M(i,j) \]
\end{defn}
\noindent Note that confusingly, the zero element for $\zinf$ is $\infty$ and the one element is $0$.  
\begin{defn}
For $[0,\infty]$-graphs $M \maps X \times X \to [0,\infty]$ and $N \maps Y \times Y \to [0,\infty]$ a function $f\maps X \to Y$ is a morphism of $[0,\infty]-graphs$ if the pushforward satisifes $f_*(M)\geq (N)$. This defines a category $[0,\infty]\Grph$ of weighted graphs and their morphisms. 
\end{defn}
\begin{prop}
There is an adjunction
\[
\begin{tikzcd}
\Set \ar[r,bend left,"L",pos=.55] \ar[r,phantom,"\bot",pos=.7]& \zinf\Grph \ar[l,bend left,"R",pos=.45]
\end{tikzcd}
\]
whose left adjoint sends a set $X$ to the $\zinf$-matrix with vertices given by $X$ and every entry given by $LX(i,j)= \infty$. The right adjoint sends a $\zinf$-graph to its underlying set of vertices.
\end{prop}
\noindent The functor $F$ in our setup will compute the shortest paths on a weighted graph. This problem can be understood algebraically. Let $\Mat(X,\zinf)$ be the semiring of $\zinf$-graphs over a set $X$. In this semiring, addition is given by pointwise minimum and multiplication is given by 
\[M \cdot N (i,j) = \min_{k \in X} \{ M(i,k) + N(k,j)\} \] which is the usual matrix multiplication valued in min-plus semiring. 
\begin{prop}
The shortest paths in a  $\zinf$-graph $M$, are given by the matrix exponential
\[ F(M) = \sum_{n \geq 0} M^n\]
in the semiring $\Mat(X,\zinf)$.
\end{prop}
\noindent This formula may be interpreted as follows: The matrix power $M^n$ has entries given by the shortest paths in exactly $n$ steps. Therefore, summing these values over all $n\geq 0$ gives the shortest paths in any number of steps.

The algebraic path problem generalizes the shortest path problem by allowing the semiring to vary. For different choices of semiring, the algebraic path problem asks for most likely paths, maximum capacities, connectivity, and more substantially the language of an NFA. As explained in \cite{master}, when $S$ is a quantale, there is an adjunction
\[
\begin{tikzcd}
S\Grph \ar[r,bend left,"F_S"] \ar[r,phantom,"\bot"]& \ar[l,bend left,"U_S",pos=.55] S\Cat 
\end{tikzcd}
\]
between the category of $S$-enriched graphs and $S$-enriched categories. The left adjoint of this adjunction sends an $S$-enriched graph to the solution of its algebraic path problem. For each left adjoint $F_S$, there there is an instance of the composition problem. In this paper, we solve the composition problem when $F_S=F$ i.e.\ the special case when $S$ is the semiring $\zinf$. Before doing this, we must state a relevant result.
\begin{prop}
 The pushout of $\zinf$-graphs is given by the pointwise sum
\[ M+_{LX} N \cong i_*(M) + j_*(N)\]
\end{prop}
\noindent In general, we use boldface to indicate the pushforward of a weighted graph along an implied function. For example, let $\bM$ and $\bN$ denote the above pushforwards of $M$ and $N$. Note that pushforward commutes with $F$ in the sense that $F(i_*(M))= i_*(F(M))$ and $F(j_*(N))=j_*(F(N))$.
\section{A Gainful Solution}
Because $F$ is a left adjoint we have reason to be optimistic about a solution to its composition problem. Left adjoints preserve pushouts so there is an isomorphism
\[F(M+_{LX} N) \cong F(M)+_{F(LX)} F(N)\]
where the pushout on the right is computed in the category of $\zinf$-enriched categories. In \cite{master}, it was shown that this pushout may be computed as
$F(U(F(M)) +_{X} U(F(N)))$. Although this gives a solution to the composition problem for $F$ it is not a practical one because the final application of $F$ is very expensive in terms of computation time. The main result of this paper is a more practical expression for this pushout.
\begin{theorem}\label{soln}
For pushouts and $F$ as above we have that 
\[F(M+_{LX} N)\cong\] \[ \sum_{n \leq |X|} \underbrace{\mathbf{F(\bM)F(\bN)F(\bM)}\ldots}_{n\text{ times}} + \underbrace{\mathbf{F(\bN)F(\bM)F(\bN)\ldots}}_{n\text{ times}}  \]
where $\mathbf{F(M)}$ and $\mathbf{F(N)}$ denote the pushforwards $i_*(F(M))$ and $j_*(F(N))$ respectively.
\end{theorem}
% \section{Fundamental Theorem}
% Suppose we have graphs $G : X \times X \to S$ and $H: Y \times Y \to S$ with an intersection. In other words, there is a pushout of functions
% \[ 
% \begin{tikzcd}
%  & X+_A Y & \\
% X \ar[ur,"i"] & &\ar[ul,"j",swap] Y \\
%  & A\ar[ul] \ar[ur]& 
% \end{tikzcd}\]
% which are the underlying sets and functions of the pushout of $S$-graphs 

% \[
% \begin{tikzcd}
%  & M+_{LX} N & \\
% M \ar[ur] & &\ar[ul] N \\
%  & LX \ar[ul] \ar[ur]& \\
% \end{tikzcd}
% \]
% then $G$ and $H$ have pushforward matrices $i_*(G) \maps X+_A Y \to S$ and $j_*(H) \maps X+_A Y \to S$.
 
\begin{proof}
Because $M+_{LX} N = \bM + \bN$ we have that 
\begin{align}\label{binomial}
    F(G+_K H) & = \sum_{n \geq 0} (G+_K H)^n \nonumber\\
    & \cong \sum_{n\geq0}(\bM + \bN)^n \nonumber\\
    & = \sum_{n \geq 0} \sum_{\Vec{v}\in \mathbf{2}^n} X_{v_1} \ldots X_{v_n} 
\end{align}
where 
\[X_{v_i} = \begin{cases} \bM &\text{ if } v_i= 0 \\ \bN & \text{ if } v_i =1 
\end{cases}
\]
and $\mathbf{2}^n$ is the set of boolean vectors $\vec{v}=(v_1,v_2,\ldots,v_n)$ with length $n$.
The last equality is true in any semiring and is a well-known as the generalization of the binomial theorem for non-commutative elements.
We define a function $\gamma \maps \sum_{n \geq 0} \mathbf{2}^n \to \mathbb{N}$ where sum now indicates the coproduct of sets. $\gamma(\vec{v})$ is called the crossing number of $\vec{v}$ and it is equal to the number of times $\vec{v}$ switches between $0$ and $1$. It may be defined by induction on the vector length i.e.$\gamma(\vec{v})=0 \text{ if } length(\vec{v})=0$ or $1$ and $\gamma((v_1,v_2,\ldots,v_n))=$
\[\gamma((v_1,v_2,\ldots,v_{n-1})) + \begin{cases}1 & \text{ if } v_n=v_{n-1}\\
0 & \text{ if } v_n \neq v_{n-1}
\end{cases}\]
The sum of Expression \ref{binomial} may be repartitioned using the crossing numbers to obtain
\begin{equation}\label{crossed}
= \sum_{n\geq 0}\resizebox{1.35\width}{1.3\height}{[} \underbrace{\sum_{i\geq 1} \bM^i \sum_{i \geq 1} \bN^i \ldots}_{n \text{ times}} + \underbrace{\sum_{i\geq 1} \bN^i \sum_{i \geq 1} \bM^i \ldots}_{n \text{ times}}\resizebox{1.35\width}{1.3\height}{]} 
\end{equation}
The last equality accounts for all terms with crossing number $n$. They may either start with $\bM$ or $\bN$ and then continue for any nonzero number of terms. Note that the maximum crossing number which may contribute to this sum is $|X|$. This is because a shortest path with crossing number greater than $|X|$ would cross at least one vertex in $|X|$ more than once and could therefore be shortened by removing a loop. Therefore Expression \ref{crossed} is equal to
\begin{equation}\label{withoutextra}
= \sum_{n \leq  |X|}\resizebox{1.35\width}{1.3\height}{[} \underbrace{\sum_{i\geq 1} \bM^i \sum_{i \geq 1} \bN^i \ldots}_{n \text{ times}} + \underbrace{\sum_{i\geq 1} \bN^i \sum_{i \geq 1} \bM^i \ldots}_{n \text{ times}}\resizebox{1.35\width}{1.3\height}{]} 
\end{equation}
The difference between the above sum and the desired expression is that each $\mathbf{F(M)}$ and $\mathbf{F(N)}$ include paths of length $0$. This causes each term of the desired result to also include terms with lower crossing number. However, because $\zinf$ is idempotent adding these terms twice does not affect the sum.
\end{proof}
% Now suppose that $S$ is the min-plus semiring $([0,\infty], \min, +)$ so that $F(M)$ computes shortest paths. In this case $F(M)$ simplifies for finite matrices.
% \begin{prop}\label{idempotence}
% Let $M :X \times X \to [0,\infty]$ be a matrix with $|X|=k$. Then
% \[F(M) = \sum_{n \geq 0} M^n = \sum_{n \leq k} M^n\]
% \end{prop}
% \begin{proof}
% The first equation follows from considering $M^{j}$ as the lengths of the shortest paths in $M$ of length $j$, and $M^k$ as the lengths of the shortest paths in $M$ of length $k$. Because $M$ only has $k$ vertices, every path with length $j>k$ must repeat a vertex and contain a loop as a subpath. These paths cannot be the shortest, because they may be made shorter by deleting the loop. Therefore, in the sum $\sum_{n \geq 0} M^n$, only the first $k$ terms contribute to the minimum.
% \end{proof}
\noindent This theorem gives an algorithm for the composition problem for $F$: simply plug $\mathbf{F(M)}$ and $\mathbf{F(N)}$ into the isomorphism of Theorem \ref{soln}. Next we use this isomorphism to find single source single target shortest paths.
\section{A Compositional Algorithm}
\begin{figure}\label{times}
    \centering
    \includegraphics[scale=.53]{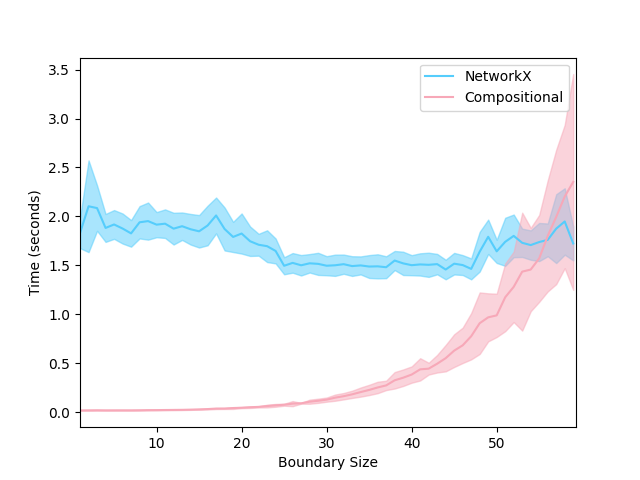}
    \caption{Comparison of Algorithms}
    \label{fig:my_label}
\end{figure}

$\mathbf{F(M)}$ and $\mathbf{F(N)}$ may be broken into the block matrices
\[\mathbf{F(M)} = \begin{bmatrix} MM & MX & 0 \\
XM & XX_{M} & 0 \\
0 & 0 & 0
\end{bmatrix}
\mathbf{F(N)}=
\begin{bmatrix}
0 & 0 & 0 \\
0 & XX_{N} & XN \\
0 & NX & NN
\end{bmatrix}\]
so that each block is labeled by the the edges that it contains. Explicitly, $MM$ consists of the edges going from $M$ to $M$, $MX$ from $M$ to $X$, $XM$ from $X$ to $M$, $XX_M$ from $X$ to $X$ within $M$ and similarly for the blocks of $\mathbf{F(N)}$. Taking the blocks as their own variables, we plug $\mathbf{F(M)}$ and $\mathbf{F(N)}$ into the isomorphism of Theorem \ref{soln} to get the \define{composition symbol matrices}: $\mathsf{Symbol(k)}= F(M+_{LX} N))$ when $|X|=k$. The \define{composition symbols}, $\mathsf{Symbol(k,i,j))}$, are the entries of these matrices. For example,
\[
\mathsf{Symbol}(4,1,3) = MX \cdot XN + MX \cdot XX_N \cdot XX_M \cdot XN
\]
The terms of $\mathsf{Symbol}(4,1,3)$ represent the paths of length less than $4$ which start in $M$ and end in $N$. The second term of this symbol represents the paths which travel between components as drawn below
\[
\begin{tikzpicture}[scale=0.3,blend group=screen]
\begin{pgfonlayer}{background}

\end{pgfonlayer}

	\begin{pgfonlayer}{nodelayer}
\node[circle,
draw=cyan,minimum size=3cm, text=black] (c) at (-3.5,0) {};
\node[circle,
draw=magenta,minimum size=3cm, text=black] (c) at (3.5,0) {};
		\node [style=dot] (0) at (-5, 3) {};
		\node [style=dot] (1) at (0, 3) {};
		\node [style=dot] (3) at (0, 0) {};
		\node [style=dot] (5) at (0, -3) {};
		\node [style=dot] (6) at (5, -3) {};
	\end{pgfonlayer}
	\begin{pgfonlayer}{edgelayer}
		\draw (0.center) -- coordinate[midway] (MX) (1.center);
		\draw [bend left=90, looseness=5.00] (1.center) to (3.center);
		\draw [bend right=90, looseness=5.00] (3.center) to (5.center);
		\draw (5.center) --coordinate[midway] (XN) (6.center);
		\node [above] at (MX) {$MX$};
			\node [right] at (4,1.5) {$XX_N$};
			\node [left] at (-4,-1.5) {$XX_M$};
			\node [below] at (XN) {$XN$};
	\end{pgfonlayer} 
\end{tikzpicture}
\]
The single source single target shortest path algorithm has three steps:
\begin{enumerate}
\item The composition symbols are computed up to the size of the boundary. This needs to be done only once for each boundary size.
    \item Precompilation. In this step, the all pairs shortest paths $F(M)$ and $F(N)$ are computed, pushed forward to $\mathbf{F(M)}$ and $\mathbf{F(N)}$, and broken into blocks as shown above.  The computations in this step only need to be done once for each input matrix so their results may be reused in all further computations.
    \item Composition. In this step, the source and target nodes $s$ and $t$ are located within the blocks of $F(M)$ and $F(N)$. The appropriate composition symbol is looked up and evaluated on the blocks of $F(M)$ and $F(N)$ using the operations of the min-plus matrix semiring. The first term of this expression is replaced by the row-vector corresponding to $s$ an the last term must be replaced by the column vector corresponding to $t$.
\end{enumerate}
An implementation in Python for this algorithm may be found at \cite{pathcomposer}. Figure \ref{times} compares this algorithm to the networkx implementation of Dijkstra's algorithm. Figure \ref{times} was computed using the pushout of two randomly weighted dense graphs with 500 nodes each. For each boundary size, the average computation time for 50 runs of both algorithms with randomly chosen source and target is plotted with the standard deviation of represented by shading. The precompilation time for this plot was 1.34 seconds.
 As the size of the boundary increases, the computation time for the compositional algorithm increases because the composition symbols grow very large. On the other hand, computation time for Dijkstra's algorithm decreases slightly as the boundary size increases because identifying more nodes reduces the total size of the graph. The speed-up from using the compositional algorithm is most dramatic when the graph size is large and the boundary size is small. For the composition of two random graphs with 2000 nodes along a boundary of size 5, the times for the compositional algorithm were sampled 50 times for a mean value of 0.1602 seconds and with standard deviation 0.0169. Dijkstra's algorithm was sampled with the same parameters for a mean value of 39.7804 seconds and standard deviation 3.3561. Regardless, there is no free lunch, precompilation for the compositional algorithm took 93.0590 seconds.
\section{Conclusion}
In this paper we found a formula for composing shortest paths and described an algorithm which implements this formula. We hope that the results and algorithm may be generalized to other instances of the algebraic path problem. In particular we are interested in the case of semiring when the algebraic path problem asks for the language of a nondeterministic finite automaton. Orthogonally, we believe the results of this paper may be extended to more complicated decompositions of weighted graphs and plan on addressing this in future work.
\bibliographystyle{alpha}

\section{Acknowledgements}
Thank you to Benjamin Bumpus, Jules Hedges, Dario Stein, Emily Roff, and Matteo Capucci who all contributed to this work.
% ADD THE FOLLOWING COUPLE LINES INTO YOUR PREAMBLE

\end{document}

%% file: preamble.tex
\usepackage{amssymb,mathrsfs,amsmath,amsfonts,graphicx,pdfpages,csquotes,amsthm}

\usepackage{lmodern}
\everymath{\fam=0\relax}

\newtheorem{theorem}{Theorem}[section]
\newtheorem{defn}{Definition}[section]
\newtheorem{prop}{Proposition}[section]

\newcommand{\bM}{\mathbf{M}}
\newcommand{\bN}{\mathbf{N}}

\newcommand{\define}[1]{{\bf \boldmath{#1}}}

\newcommand{\namedcat}[1]{\mathsf{#1}}

\newcommand{\Cat}{\namedcat{Cat}}

\newcommand{\Set}{\namedcat{Set}}

\newcommand{\Mat}{\mathsf{Mat}}

\newcommand{\Grph}{\mathsf{Graph}}
\newcommand{\zinf}{[0,\infty]}
\makeatletter
\newcommand*{\relrelbarsep}{.386ex}
\newcommand*{\relrelbar}{%
  \mathrel{%
    \mathpalette\@relrelbar\relrelbarsep
  }%
}
\newcommand*{\@relrelbar}[2]{%
  \raise#2\hbox to 0pt{$\m@th#1\relbar$\hss}%
  \lower#2\hbox{$\m@th#1\relbar$}%
}
\providecommand*{\rightrightarrowsfill@}{%
  \arrowfill@\relrelbar\relrelbar\rightrightarrows
}
\providecommand*{\leftleftarrowsfill@}{%
  \arrowfill@\leftleftarrows\relrelbar\relrelbar
}
\providecommand*{\xrightrightarrows}[2][]{%
  \ext@arrow 0359\rightrightarrowsfill@{#1}{#2}%
}
\providecommand*{\xleftleftarrows}[2][]{%
  \ext@arrow 3095\leftleftarrowsfill@{#1}{#2}%
}
\makeatother

\usepackage{subfiles}
\usepackage{stackengine}
\usepackage{mathtools}
\usepackage[utf8]{inputenc}
\usepackage{tikz}

\usepackage{comment}

\usepackage{graphicx}
\usepackage{adjustbox}
\usepackage[all,2cell]{xy}\UseAllTwocells\SilentMatrices
\definecolor{darkgreen}{rgb}{0,0.45,0}
\definecolor{myurlcolor}{rgb}{0,.45,.2}
\definecolor{mylinkcolor}{rgb}{.1,.1,.7}
\usepackage[colorlinks,citecolor=darkgreen,urlcolor=myurlcolor,linkcolor=mylinkcolor,final,hyperindex,linktoc=page,pagebackref]{hyperref}

\usepackage[capitalize]{cleveref}
\crefname{equation}{}{}
\crefname{item}{}{}
\usepackage{enumerate}

\usepackage{titlesec}
\titlespacing*{\section}{0pt}{0.2\baselineskip}{0.2\baselineskip}
\titleformat{\section}
  {\normalfont\fontsize{12}{15}\bfseries}{\thesection}{1em}{}
\newcommand{\maps}{\colon}

\let\OLDthebibliography\thebibliography
\renewcommand\thebibliography[1]{
  \OLDthebibliography{#1}
  \setlength{\parskip}{0pt}
  \setlength{\itemsep}{0pt plus 0.3ex}
}
\usepackage{tikz}
\usepackage{tikz-cd}
\usetikzlibrary{backgrounds,circuits,circuits.ee.IEC,shapes,fit,matrix}
\usepackage{verbatim}
\usetikzlibrary{arrows,shapes}

\tikzstyle{dot}=[circle,fill=black]

\tikzstyle{simple}=[-,line width=2.000]
\tikzstyle{arrow}=[-,postaction={decorate},decoration={markings,mark=at position .5 with {\arrow{>}}},line width=1.100]
\pgfdeclarelayer{edgelayer}
\pgfdeclarelayer{nodelayer}
\pgfdeclarelayer{background}
\pgfsetlayers{edgelayer,nodelayer,main,background}

\tikzstyle{none}=[inner sep=0pt]

\definecolor{lblue}{rgb}{0,250,255}
\tikzstyle{species}=[circle,fill=yellow,draw=black,scale=1.15]
\tikzstyle{transition}=[rectangle,fill=lblue,draw=black,scale=1.15]
\tikzstyle{inarrow}=[->, >=stealth, shorten >=.03cm,line width=1.5]
\tikzstyle{empty}=[circle,fill=none, draw=none]
\tikzstyle{inputdot}=[circle,fill=black,draw=black, scale=.25]
\tikzstyle{inputarrow}=[->,draw=purple, shorten >=.05cm]
\tikzstyle{simple}=[-,draw=black,line width=1.000]
\tikzstyle{dot}=[circle,fill=black,draw=black, scale=.4]
\tikzstyle{inarrow}=[->, >=stealth, shorten >=.03cm,line width=1.5]

\definecolor{joecolor(x11)}{rgb}{0.0, 0.5, 0.5}

\definecolor{purple(x11)}{rgb}{0.8, 0, 0.8}